\documentclass[prl,amsmath,amssymb,floatfix,twocolumn]{revtex4}
\usepackage{graphicx}
\usepackage{float}
\usepackage{bm}
\usepackage[utf8]{inputenc}

\begin{document}

\title{Closed-loop liquid-vapor equilibrium in a one-component system}
\author{N. G. Almarza}
\affiliation{Instituto de Qu{\'\i}mica F{\'\i}sica Rocasolano, CSIC, Serrano 119, E-28006 Madrid, Spain }

\begin{abstract}
We report Monte Carlo simulations that show closed-loop liquid-vapor 
equilibrium in a pure substance.  
As far as we know, this is the first time that such a topology
of the phase diagram has been found for  one-component systems.

This finding has been achieved on a two-dimensional lattice model for  
patchy particles that can form network fluids. 
We have  considered related models with a slightly different patch 
distribution in the order to understand the features of the distribution 
of patches on the surface of the particles that make possible the presence 
of the closed-loop liquid-vapor equilibrium, and its relation
with the phase diagram containing the so-called empty liquids.  

Finally we discuss the likelihood of finding the closed-loop liquid-vapor 
equilibria on related models for three dimensional models
of patchy particles in the continuum, and speculate on the possible relationship
between the mechanism behind the closed-loop liquid vapor equilibrium  of our 
simple lattice model and
the salt-induced reentrant condensation found in complex systems.
\end{abstract}
\date{\today}
\maketitle
The recent advances in the chemical synthesis of colloidal particles 
monodisperse in shape and size have attracted a lot of theoretical
work to understand their collective behavior.
\cite{Pawar,Sciortino}. 
These new particles can be thought as `colloidal molecules'  than can 
organize themselves by self-assembly into useful structures\cite{Glotzer}. 
Patchy colloids\cite{Bianchi} are very interesting because
of the possibility  of tuning the number, interaction, and local 
arrangements of their patches.
The structure of the self-assembled clusters of the patchy colloids can 
produce therefore novel macroscopic behavior.
\cite{Pawar,Sciortino,Glotzer,Bianchi}
On the other hand, primitive models of patchy particles have been used  to represent the specific
anisotropic interactions between globular proteins in solution\cite{Bianchi}.
These systems present sometimes reentrant condensation induced by variations
of salt concentration.\cite{ZhangPRL,ZhangProteins} This behavior is probably associated to effective
charge inversion of proteins in the reentrant regime via binding of multivalent
counterions\cite{ZhangProteins}. The mechanisms involved in the condensation and
the dependence of the reentrant behavior with the temperature are not well understood,
but it is likely that  specific (patchy-like) interactions\cite{Piazza,ZhangFD} play
an important role.

Given the complexity of both kinds of problems: self-assembling of patchy colloids, 
and phase behavior of protein solutions, it seems important the study of simple models
of patchy particles, in order to understand the influence of the numbers of patches
and their distribution on the surface of the macromolecules (or colloidal particles) on
the phase behavior and aggregation mechanisms. Hopefully, understanding
the phase behavior of
simple models will be useful to gain insight in the behavior of those complex systems.

Recently it has been found that patchy models with two types of patches, 
$A$, and $B$ in which only $A-A$, and $A-B$ interactions are attractive 
can exhibit liquid-vapor equilibria (LVE) with liquid phases behaving as  `empty liquids' \cite{Russo1,Russo2,2A2B}.
 For these systems the term empty liquid\cite{Bianchi} refers to the fact that
at low temperatures the density of the liquid phase, $\rho_L$ on the binodal becomes
very small at low.  Actually  $\rho_L$ approaches zero in the limit $T\rightarrow 0$.
The empty liquid phenomenology has been found for hard 
sphere particles that carry on the surface two patches of type $A$ located 
at opposite poles of the sphere, and several patches of type $B$ elsewhere 
on the surface.  The potential energy is defined as a sum of bonding interactions
between pairs of patches belonging to different particles.

Within  this interaction scheme, and assuming for simplicity that the 
interaction between two particles is zero if neither and $AB$ nor an $AA$ 
bond between them is formed, the phase diagram basically depends on the 
ratio between the bonding energies of the two types of possible bonds: 
$y = \epsilon_{AB}/\epsilon_{AA}$.
From both Monte Carlo Simulation and Wertheim's perturbation 
theory\cite{Russo1,Russo2,2A2B,Tavares2009A} it is known that the regime 
of empty liquid may occur  for $y < 1/2$. 
Theoretical approximations 
predict\cite{Tavares2009A} also a lower limit of this ratio, 
namely $y_l=1/3$. Simulation results are consistent
with the existence of such a lower bound $(y_l>0)$ for the empty liquid regime.
On the other hand, for $y=0$, the systems at low temperature are expected to
form linear chains\cite{Giacometti}, which, may give rise to orientational 
ordered (nematic) phases at moderate densities.\cite{SARR,TSARR,Reply}  
This orientational order can also appear as a second phase
transition for values of $y<0.5$  where the liquid-vapor is present.
The topology of the empty liquid phases is well understood.
\cite{Russo1,Russo2,2A2B,Tavares2009A} At low temperature 
(where the empty liquid regime occurs) and not too low density almost all 
the $A$-patches are participating in a bond, the liquid phases can be 
then described as a network fluid composed by (almost) straight chains 
of spheres linked by sequences of $AA$ bonds.
Each of the terminal spheres of the chains is connected to another straight 
chain,  by an $AB$ bond.
In the empty liquid regime the reduction of the temperature implies
an increase of the typical length of the $AA$ chains, which in turn, 
produces a reduction of the density of the liquid phase.
This is due to the fact that $AA$ bonds  for $y<1/2$.

At this point, we can ask ourselves to what extent this physical picture 
remains valid if the nature of the colloid surface is changed, and the 
two $A$-patches are no longer located at opposite poles, but somewhat closer. 
We can define a bending angle, $\theta_{ijk}$ for a triplet of particles, 
$\{i,j,k\}$ in which $j$ is linked through $AA$ bonds
to both $i$, and $k$, as:
\begin{equation}
\cos \theta_{ijk} = \frac{ \vec{r}_{ji}  \cdot \vec{r}_{jk} }
{|\vec{r}_{ji} | |\vec{r}_{jk} | } .
\end{equation}
 Taking into account that the size of the patches is supposed to be small
in order to preserve the conditions of only one bond per patch\cite{Russo1}, 
the mean value of $\cos \theta_{ijk}$  will fulfill: 
$\langle \cos \theta_{ijk} \rangle \simeq \cos \alpha_{AA}$,
where $\alpha_{AA}$ is the angle between the vectors that connect the 
center of the sphere with the centers of its $A$ patches.  
Notice that reducing  the value of $\alpha_{AA}$ from $180^{\circ}$ 
implies an increase the probability of forming rings of particles connected 
by $AA$-bonds, since the system can reduce its energy by forming cycles. 
Moreover, when one of these cycles is formed all the $A$-patches of
its particles are bonded, and therefore these patches cannot bond other 
chains, which in turn can reduce the connectivity of different parts of 
the system, and eventually inhibit the appearance of LVE.

Now taking into account that for $\alpha_{AA}=180^{\circ}$ and appropriate 
values of $y$ one can find the empty liquid regime at low temperatures, 
and that reducing the value of $\alpha_{AA}$ might, in principle, destroy 
the LVE, it seems interesting to know what happens between the two scenarios. 
In order to investigate this issue, we have found convenient to analyze a 
simple lattice model whose results are expected to be representative of what 
can be found for models in the continuum, as it has be shown
to occur in related models\cite{2A2B,Noya}. The computation of phase diagrams
is usually much cheaper in terms of CPU time in lattice models than in models 
defined on the continuum. In spite of their simplicity,
lattice models can often capture the essential features of the phase diagrams 
of associating molecules and network fluids
\cite{LRampa3d,LRampa2d,2A2B,Noya,Roberts,Szortyka1,Szortyka2,Szortyka3,
MenonEPL,MenonPRE}.

\begin{figure}[ht]
\includegraphics[width=90mm,clip=]{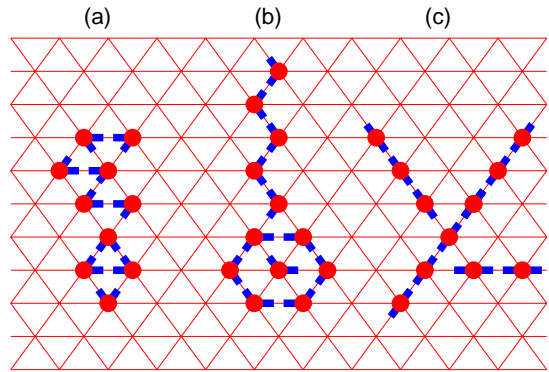}
\caption{Sketch of the patchy particles in the three lattice models: 
(a) {\it o-}2A4B model: $\alpha_{AA}=60^{\circ}$;
(b) {\it m-}2A4B model: $\alpha_{AA}=120^{\circ}$;
(c) {\it p-}2A4B model: $\alpha_{AA}= 180^{\circ}$. Thin lines
represent the underlying triangular lattice. Filled circles represent
particles (occupied sites). Thick segments indicate the orientation
of $A$-patches.}
\label{FIG1}
\end{figure}

We have considered models on the triangular lattice 
(See Fig. \ref{FIG1}).  
Sites on the lattice can be either empty or occupied by one particle. 
Each particle carries two sites (of type A)  that point
to two of the nearest neighbor (NN) sites. 
Two particles, $i,j$; being NN interact with
an energy $u_{ij} = -\epsilon$ if each one points to the other with 
one of its $A$-patches ($AA$ bond);
if only one of the particles points to the other with one of its $A$ patches, 
we have $u_{ij} = -y \epsilon$ ($AB$ bond, 
with $0<y\le 1/2$); and otherwise $u_{ij}=0$.

 We have studied  three models (See FIG. \ref{FIG1}), 
which will be denominated as {\it o-}, {\it m-} and {\it p}-2A4B models 
(by borrowing the prefixes from the organic chemistry nomenclature for  
the positions of the substituents in aromatic cyclic compounds: ortho, 
meta and para). The three models differ in the arrangements of the
patches. In {\it o-}2A4B  we have $\alpha_{AA}= 60 ^{\circ}$,
for {\it m-}2A4B $\alpha_{AA} = 120^{\circ}$, and finally {\it p-}2A4B model
considers $\alpha_{AA}=180^{\circ}$. Notice that the latest model is the natural
translation of previous works for empty fluids into the triangular lattice.
In the {\it o-}2A4B model cycles of $AA$ bonds are easily formed, the smallest
possible cycle contains just three particles, in the {\it m-}2A4B model we can
find cycles of six $AA$-bonds, whereas in the {\it p-}2A4B models no cycles
are possible.
In order to compute the phase diagram we have made use of simulation strategies
that were described in previous papers\cite{2A2B,Noya,LRampa2d,LRampa3d}; 
namely Wang-Landau multicanonical simulations\cite{2A2B,Lomba}  to locate critical points 
and some liquid-vapor coexistence points;
and Gibbs-Duhem integration to trace the binodals. Some special techniques 
have been developed in the simulation procedures to improve the sampling, 
including efficient cluster algorithms for the {\it p-}2A4B model. 
Technical details will be published elsewhere \cite{tobepublished}

\begin{figure}[h]
\includegraphics[width=80mm,clip=]{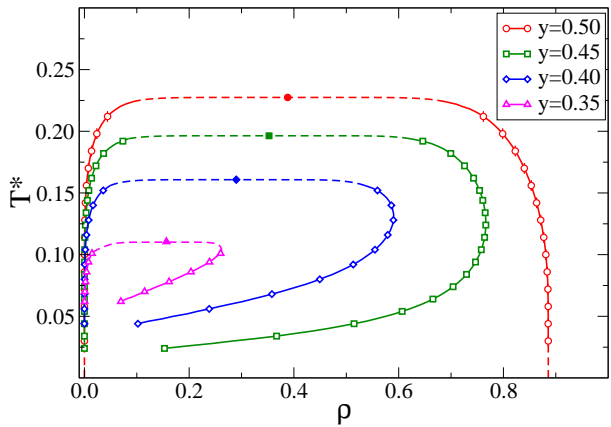}
\caption{Temperature-density phase diagrams for the {\it p-}2A4B model, and
different values of $y$ (See the legends).
Filled symbols indicate critical points. $T^*$ is the reduced temperature:
$T^*=k_B T/\epsilon$, and the density $\rho$ corresponds to the fraction
of occupied sites.}
\label{FIG2}
\end{figure}

We have  have not found evidence of LVE for the {\it o-}2A4B model. 
This seems plausible, since at low temperature and moderate density 
the particles can easily arrange themselves in triplets that minimize
the energy and block the development of an extended network of patch-patch 
bonds.
Both {\it m-}2A4B and {\it p-}2A4B models show LVE for some range
of values in the region $y<1/2$.  The {\it p-}2A4B model
shows, as expected, a phase diagram (See FIG \ref{FIG2}) that fully 
resembles those of the  analogue models in the continuum \cite{Russo1,Russo2}, 
and on the square lattice \cite{2A2B};
i.e. it appears the regime of empty liquid. 

\begin{figure}[ht]
\includegraphics[width=80mm,clip=]{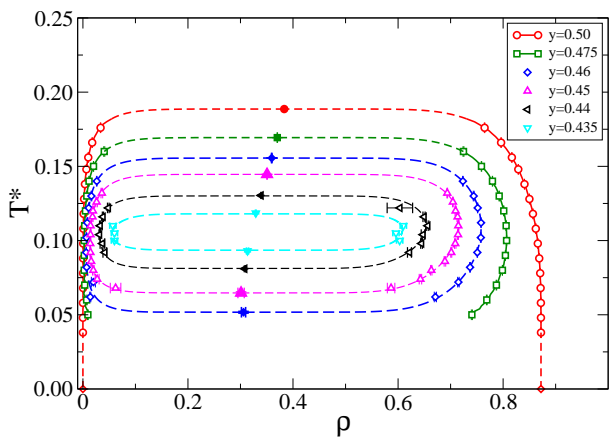}
\caption{Temperature-density phase diagrams for {\it m-}2A4B model. 
See FIG. \ref{FIG2} caption for notation.}
\label{FIG3}
\end{figure}

The key result of this paper is, however,  the phase
diagram of the {\it m-}2A4B model (shown in Fig \ref{FIG3}). 
 For $y=0.50$ we get a similar behavior in {\it m-}2A4B and {\it p-}2A4B 
models, i.e, we find typical LVE binodals.
However, in the {\it m-}2A4B model we find that the LVE 
for some range of values of $y$ ( $y_l < y < 1/2)$, with $y_l \simeq 0.43$, 
exhibits two critical points, that will be named, according to their critical 
temperature as  lower critical point (LCP) and upper critical point (UCP).
The LVE only occurs between these two temperatures, which of course
depend on the value of $y$. The critical points were determined through 
a finite-size scaling analysis 
on the multicanonical simulation
similar to that explained in Ref. \onlinecite{2A2B}
 considering typically system sizes in the
range $12 \le L \le 48$ for the {\it m-}2A-4B model (with $L^2$ being the
total number of sites). The scaling behavior was found to be fully consistent
with the two-dimensional Ising criticality for both LCP and UCP.

The existence of reentrant phase equilibria  is well
known to appear in some binary mixtures.  
However, for 
one-component systems it is very unusual to find fluid-fluid reentrant 
critical behavior.
Roberts et al. \cite{Roberts} found such a behavior for a lattice model 
of network-forming fluids:  However the closed-loop coexistence found 
in Ref. [\onlinecite{Roberts}] involves liquid phases.
In our case, the phase equilibria can be appropriately described as LVE, by considering
that modifying the value of the interaction parameter $y$
the reentrant phase diagrams can be  connected continuously to typical LVE diagrams that
appear for  $y\ge1/2$.
Along this path the  LVE envelope varies smoothly for $T>0$.

\begin{figure}[ht]
\includegraphics[width=80mm,clip=]{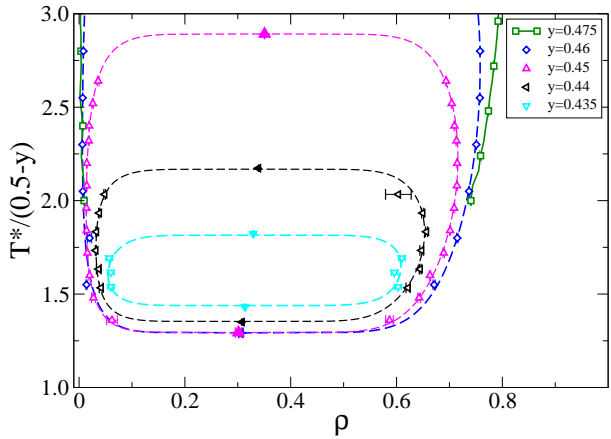}
\caption{Liquid vapor equilibria phase diagrams for {\it m-}2A4B  model with
an alternative temperature scaling.}
\label{FIG4}
\end{figure}

In the previous paragraph we have indicated  that the closed-loop LVE
occurs for any value of $y$ in the range $(y_l,1/2)$, however  our 
simulation results only support, in principle, the reentrance for 
$y_l < y  \le  0.46$.  As $y$ approaches
1/2 it becomes a very hard task to compute the LCP using Monte Carlo 
simulation, since it occurs at very low temperatures. 
However, we can make use of the scaling criteria introduced in 
Ref.\onlinecite{2A2B}:
At very low temperature, the probability of finding non-bonded $A$ patches 
becomes negligible small, therefore we can consider as a good approximation 
to analyze  the phase equilibria  considering only those configurations in which each 
patch $A$ forms either an $AA$ or an $AB$ bond. 
Under this assumption the potential energy of the model can be written 
as:\cite{2A2B} 
$(U/\epsilon)  = - N + {\cal N}_{AB} \left( \frac{1}{2} - y \right)$,
where $N$ is the number of occupied sites and ${\cal N}_{AB}$ the number 
of $AB$ bonds,
Therefore  the  phase behavior at low temperature, $T$, will basically 
depend on the ratio $(1/2-y)\epsilon/k_BT$ (with $k_B$ being the 
Boltzmann constant). Then the
temperature at the LCP is expected to scale for $(1/2-y) \rightarrow 0^+$ 
as:
$T_{C}^{l} \propto (1/2-y)$. 
The low temperature phase diagram of the
{\it m}-2A4B model has been found to fulfill very well this
scaling at low temperature (See FIG. \ref{FIG4}). 
Of course the reliability of this prediction depends
on the absence of additional possible phase transitions that could occur 
in the model.
Notice that the {\it p-}2A4B model is expected to exhibit and order-disorder 
transition.\cite{2A2B,SARR,TSARR,Reply}
Such a transition will merge with the LV transition as $T\rightarrow 0$.
In the case of the {\it m-}2A4B model we have not found any signature 
of a second phase transition at the temperature range where the close-loop 
LVE occurs.
\begin{figure}
\includegraphics[width=85mm,clip=]{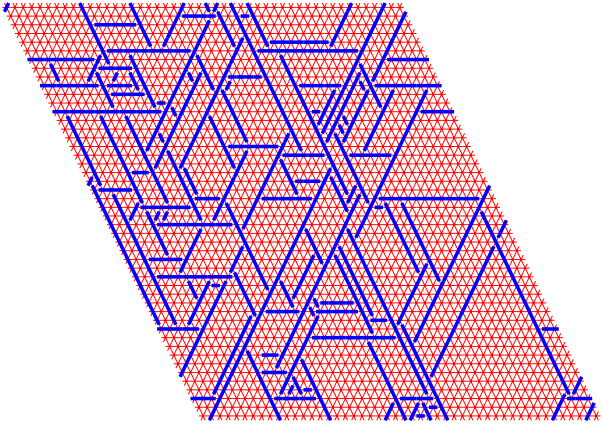}
\caption{Configuration for the liquid phase at the binodal for the 
{\it p}-2A4B model with $y=0.40$,  $T*=0.07$, $\rho \simeq 0.375$}
\label{FIG5}
\end{figure}
\begin{figure}
\includegraphics[width=85mm,clip=]{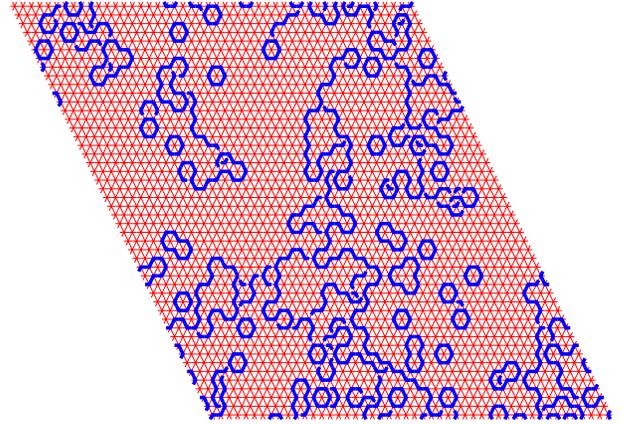}
\caption{Configuration for the {\it m}-2A4B model with $\rho = 1/3$, $T*=0.03$
(below the lower critical temperature).}
\label{FIG6}
\end{figure}
\begin{figure}
\includegraphics[width=85mm,clip=]{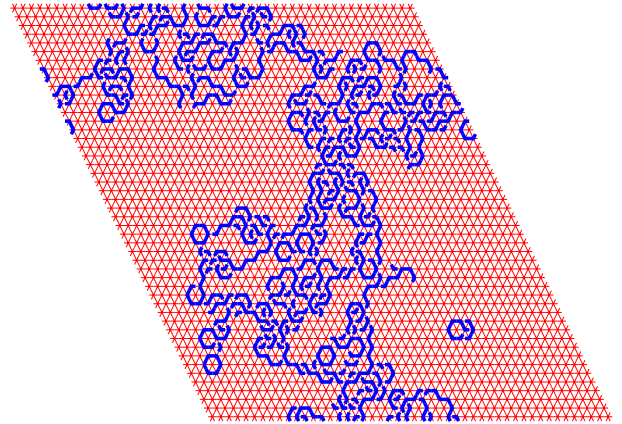}
\caption{Configuration for the {\it m}-2A4B model with $\rho = 1/3$, $T*=0.10$.
At these conditions there is LV phase separation}
\label{FIG7}
\end{figure}
The origin of the different topologies of the LVE phase diagrams 
of {\it p}- and {\it m}-2A4B models can be explained by looking at 
representative configurations of both models at low temperature 
(FIGS \ref{FIG5}-\ref{FIG7}). 
In the {\it p}-2A4B model the end particles of each chain are bonded 
via $AB$-bonds to two other chains on the system, then
on average each chain is bonded to four neighbor chains 
(${\cal N}_{AB} = 2 N_{\textrm chains}$ at low T). 
At very low temperature in the {\it m}-2A4B model we can see
(See FIG \ref{FIG6})  the formation of rings with different number of particles, and the existence
of AB intra-chain bonds. As a consequence the number of available  $A$ patches to form interchain
bonds is reduced, and the bonds do not  produce a macroscopic bond network. Notice
that as one increases the temperature (See FIG \ref{FIG7}) there is a reduction of the number of
 rings
due to entropic effects, \cite{Tavares2012} this enhances the connectivity and the phase separation can take place,
by a mechanism similar to that present in the {\it p}-2A4B model.

 It has to be emphasized that the origin of the reentrance of the LVE is not just
the angle between patches in the colloidal particles, but the fact that at low temperature
the particles have the trend to
self-assemble forming small aggregates (usually rings) in which most of their sticky spots
are saturated by intra-aggregate interactions. This process might produce
a reduction of the {\it effective valence} of the aggregates with respect to that
of the independent particles, which in turns reduces the possibility of bond
network formation and subsequently hinders the ability of the system to 
condense.

The mechanism underlying the reentrance of the LVE is rather different to that 
appearing in binary liquid mixtures, in which the reentrant solubility at low
temperature is attributed to the effects of large anisotropic interactions (hydrogen bonding)
between the two components, whereas the demixing at intermediate temperatures
is due to Wan der Waals interactions, which favor the interaction between pairs of particles
of the same species.\cite{Hirschfelder,Goldstein}

Notice that the results presented in this communication resemble, 
to some extent, those reported by Sciortino et al.\cite{PRL103,PCCP} for 
a system of {\it Janus} particles, 
 and by Reinhardt {\it et al.}\cite{Reinhardt} for systems with competition between
phase separation and self-assembly.
In both cases,
 at low temperature,
the vapor density at LVE increases on cooling. However in these systems no LCP appears,
due to the presence of crystalline phases  that make both liquid and vapor 
metastable prior the possible presence of the LCP.

The question that arises after having found the closed-loop LVE in the 
lattice model for systems with two A-patches with $\alpha_{AA} = 120^{\circ}$, 
is to what extent these results can be extrapolated to models in the 
continuum, either in two or three dimensions.
In principle, one can expect that for 2D-models the formation of rings, 
which in turn may produce the closed-loop LVE, will be enhanced for models 
with angles $\alpha_{AA}$ that fit those of regular polygons 
$\alpha_{AA} \simeq \frac{n-2}{n} \pi $, with $n=4, 5, 6, \cdots$.
For 3D models one can also expect the formation of rings, in this case 
the presence of additional  degrees of freedom in the chains will give
the systems more {\it flexibility} for the formation of rings,
however, on the other side,  the larger space dimensionality of the system 
will favor the presence of open chains due to  entropy effects.
In any case the pursue of the conditions whether closed -loop
LVE will be obviously not so straightforward as in the current lattice models.
We plan to work on these issues in the future.

As a summary, we have shown how patchy models on the triangular lattice 
can exhibit closed-loop LVE on one-component systems. The basic physics 
underlying this behavior is the reduction of valence at low temperature 
by the formation of compact structures at low and moderate densities that 
reduce the amount of connectivity between different parts of the systems. 
The phase diagram depends dramatically on the
angle between the $A$ patches. As far as we know, the work presented here
constitutes the first model exhibiting closed-loop LVE on a one-component system.
Finally, self assembling mechanisms similar to those appearing
in the simple models presented in this paper, might play some role in
the reentrant condensation phenomena of globular proteins\cite{ZhangFD} and polyelectrolytes
\cite{Hisao}.

\acknowledgments
The author gratefully acknowledges the support from the Direcci\'on General de Investigaci\'on Cient\'{\i}fica  y T\'ecnica under Grant No. FIS2010-15502, and from the
Direcci\'on General de Universidades e Investigaci\'on de la Comunidad de Madrid under Grant No. S2009/ESP-1691 and Program MODELICO-CM.

\end{document}